\documentclass[12pt]{article}
\usepackage{cite}
\usepackage[dvips]{graphicx}
\usepackage{epsfig}
\usepackage{amssymb}
\usepackage{rotating}

\begin{document}
\begin{flushright}
MAN/HEP/2008/21\\
arXiv:0808.0490\\
August 2008        
\end{flushright}

\bigskip

\begin{center}
{\LARGE\bf \hspace{-5mm}Thermal~Right-Handed~Sneutrino~Dark~Matter \\[0.3cm] 
in the {\boldmath $F_D$}-Term Model of Hybrid Inflation}\\[1.5cm]
{\large Frank Deppisch
%
and Apostolos Pilaftsis }\\[0.3cm]
{\em School of Physics and Astronomy, University of Manchester,\\ 
     Manchester M13 9PL, United Kingdom}
\end{center}

\vspace{1.cm}\centerline{\bf ABSTRACT}

\noindent
We compute the  relic abundance of the right-handed  sneutrinos in the
supersymmetric  $F_D$-term  model of  hybrid  inflation.   As well  as
providing a natural solution to the $\mu$- and gravitino overabundance
problems,  the  $F_D$-term model  offers  a  new  viable candidate  to
account  for  the cold  dark  matter  in  the Universe:  the  lightest
right-handed sneutrino.  In  particular, the $F_D$-term model predicts
a new quartic coupling of  purely right-handed sneutrinos to the Higgs
doublets  that thermalizes  the sneutrinos  and makes  them annihilate
sufficiently  fast  to a  level  compatible  with  the current  cosmic
microwave  background data.  We  analyze this  scenario in  detail and
identify  favourable  regions  of   the  parameter  space  within  the
framework of minimal supergravity, for which the lightest right-handed
sneutrino  becomes the {\em  thermal} dark  matter, in  agreement with
WMAP observations of cosmological inflation.  Constraints derived from
direct dark matter searches experiments are presented.

\noindent

\medskip
\noindent
{\small PACS numbers: 98.80.Cq, 12.60.Jv, 11.30Pb\\
{\sc Keywords:} Supersymmetry, Dark Matter, Inflation}

\newpage

\setcounter{equation}{0}
\section{Introduction}

Hybrid   inflation~\cite{Linde},   along   with   its   supersymmetric
realizations~\cite{CLLSW,DSS,Halyo,CP},   remains  one  of   the  most
predictive and  potentially testable scenarios of  inflation that have
been suggested so far.  Hybrid inflation is predictive and testable, in the  sense that the inflaton dynamics  is mainly governed
by  a  few  renormalizable   operators  which  might  have  observable
implications for laboratory experiments. In such a scenario, inflation
terminates  through  the   so-called  waterfall  mechanism,  which  is
triggered, when  the inflaton field $\phi$ passes  below some critical
value  $\phi_c$. From  that point  on, another  field $X$,  called the
waterfall field, held fixed at origin initially, quickly rolls down to
its true  vacuum expectation value~(VEV) and  drastically modifies the
slow-roll form of the $\phi$-potential, thereby ending inflation.

In  supersymmetric   theories,  the   required  form  of   the  hybrid
inflationary potential may originate  from either the $F$-terms of the
superpotential     or    from    a     large    Fayet--Iliopoulos~(FI)
$D$-term~\cite{FI},  usually  induced  by  some anomalous  local  U(1)
symmetry within the context of  string theories.  In both the $F$- and
$D$-term hybrid  inflation, the slow-roll  slope of the  potential may
come either from  supergravity (SUGRA) corrections~\cite{CLLSW} and/or
from radiative effects~\cite{DSS,Halyo,CP}.

Recently, a new supersymmetric  hybrid inflationary model was proposed
in~\cite{GP} and studied in  detail in~\cite{GPP}.  The model realizes
$F$-term   hybrid   inflation   and   includes  a   subdominant   {\em
non-anomalous}  FI  $D$-term  that  arises  from  the  U(1)$_X$  gauge
symmetry of  the waterfall  sector. It has  therefore been  called the
$F_D$-term  model of  hybrid inflation,  or in  short,  the $F_D$-term
model.  The  $F_D$-term model can naturally  accommodate the currently
favoured    red-tilted   spectrum    with    $n_{\rm   s}-1    \approx
-0.037$~\cite{Dunkley:2008ie},  along  with the  actual  value of  the
power  spectrum   of  curvature  perturbations,   $P_{\cal  R}  \simeq
4.86\times   10^{-5}$~\cite{WMAP3},  and   the   required  number   of
$e$-folds, ${\cal N}_e \approx 55$~\cite{review}.

The presence  of the FI term  in the $F_D$-term model  is necessary to
approximately break  a $D$-parity  that governs the  waterfall sector.
The approximate breaking  of the $D$-parity gives rise  to late decays
of the  superheavy waterfall-sector  particles that are  produced just
after         inflation          during         the         preheating
epoch~\cite{PREHEATING,GBRM}. These waterfall particles have masses of
the  Grand Unified  Theory (GUT)  scale  and can  dominate the  energy
density of  the Universe, provided  the inflaton coupling  $\kappa$ to
the waterfall sector is not too suppressed, i.e.~for values of $\kappa
\stackrel{>}{{}_\sim}  10^{-3}$.    Then,  the  late   decays  of  the
GUT-scale  waterfall particles  produce an  enormous entropy  that can
reduce  the  gravitino abundance  $Y_{\widetilde{G}}$  well below  the
limits     imposed    by     big    bang     nucleosynthesis    (BBN),
i.e.~$Y_{\widetilde{G}}                           \stackrel{<}{{}_\sim}
10^{-15}$~\cite{gravitino}.   In   this  way,  the   $F_D$-term  model
provides   a   viable   solution   to  the   gravitino   overabundance
problem~\cite{GPP},  without  the  need  to unnaturally  suppress  all
renormalizable  inflaton couplings  to  the particles  of the  Minimal
Supersymmetric  Standard  Model  (MSSM)  sector, below  the  $10^{-6}$
level.

Another  interesting  feature of  the  $F_D$-term  model  is that  the
$\mu$-parameter  of  the MSSM  can  be  generated  effectively by  the
superpotential    operator    $\lambda    \widehat{S}    \widehat{H}_u
\widehat{H}_d$,  when  the scalar  component  of  the inflaton  chiral
multiplet $\widehat{S}$ receives a  non-zero VEV after the spontaneous
symmetry  breaking  (SSB)  of  the  local  U(1)$_X$  symmetry  of  the
waterfall   sector~\cite{DLS}.   Moreover,  the   inflaton  superfield
$\widehat{S}$     couples     to     the     right-handed     neutrino
superfields~$\widehat{N}_{1,2,3}$,  via  the  superpotential  coupling
$\frac{1}{2}\rho_{ij} \widehat{S}  \widehat{N}_i\widehat{N}_j$, with $i,j=1,2,3$.
Hence, the inflaton VEV will produce an effective Majorana mass matrix
as  well~\cite{Francesca,GP}.  As a  consequence, the  resulting heavy
Majorana neutrinos  are expected  to have masses  of order  $\mu$.  If
$\rho_{ij}$ is approximately  SO(3) symmetric, i.e.~$\rho_{ij} \approx
\rho\,  {\bf 1}_3$,  a  possible explanation  of  the observed  baryon
asymmetry  in   the  Universe  (BAU)   may  be  obtained   by  thermal
electroweak-scale resonant  leptogenesis, in a way  independent of any
pre-existing lepton- or baryon-number asymmetry~\cite{PU2}.

Even though the $F_D$-term model violates explicitly the lepton number
($L$)  by  $\Delta  L  =  2$ superpotential  operators,  it  conserves
$R$-parity.   Hence,  the  lightest  supersymmetric particle  will  be
stable and  so will  potentially qualify as  a candidate for  the cold
dark matter (DM) in  the Universe.  Most interestingly, the $F_D$-term
model provides a new candidate for  the cold DM.  This is the lightest
right-handed  sneutrino  (LRHS),   which  may  possess  thermal  relic
abundance~\cite{GPP} for relatively large values of the aforementioned
superpotential  couplings $\lambda$  and $\rho$,  i.e.~for $\lambda,\,
\rho \stackrel{>}{{}_\sim}  10^{-2}$.  This should  be contrasted with
what is  happening in  standard seesaw extensions  of the  MSSM, where
$\widehat{N}_{1,2,3}$  have only  bare Majorana  masses.   Because the
small neutrino Yukawa couplings  are the only possible interactions of
sneutrinos with matter in these models, purely right-handed sneutrinos
turn out to be non-thermal and  tend to overclose the Universe by many
orders of magnitude~\cite{GGP,LMN}.  It is therefore difficult for the
LRHS to  be a thermal  DM in seesaw  extensions of the MSSM  with bare
Majorana masses.

In  this  paper  we analyze  in  detail  the  relic abundance  of  the
right-handed  sneutrinos  in the  supersymmetric  $F_D$-term model  of
hybrid  inflation.   In  this  model,  the $F$-term  of  the  inflaton
superfield,   $F_S$,  gives   rise  to   the  new   quartic  coupling,
$\frac{1}{2}\lambda\rho\,  \widetilde{N}^*_i \widetilde{N}^*_i  H_u H_d$,  in the
scalar   potential,  which   involves   the  right-handed   sneutrinos
$\widetilde{N}_{1,2,3}$   and   the   Higgs  doublets~$H_{u,d}$.    As
mentioned  above, unless the  couplings $\lambda$  and $\rho$  are too
small,  the  new  quartic  coupling  will be  sufficiently  strong  to
thermalize  the  sneutrinos  and  make  them  annihilate  to  a  level
compatible  with   the  current  cosmic   microwave  background  (CMB)
data~\cite{Dunkley:2008ie},  from  which   the  DM  component  of  the
Universe was found to be
\begin{equation}
  \label{OmegaDM}
\Omega_{\rm   DM}\,h^2\ =\ 0.1099\: \pm\: 0.0062\; .
\end{equation}
The central goal  of our analysis is to  delineate the parameter space
within  the context of  minimal supergravity  (mSUGRA), for  which the
LRHS is the {\em thermal} DM. In addition, we consider the constraints
obtained  by  WMAP  observations  related to  cosmological  inflation.
Finally,   we   present   numerical   estimates  of   the   scattering
cross-section of the LRHS with  nuclei that will be relevant to direct
DM searches in present and future experiments.

After  this  introduction,  the  paper  is organized  as  follows:  in
Section~\ref{FDmodel} we present the basic structure of the $F_D$-term
model and  briefly review the solution to  the gravitino overabundance
problem.   Moreover, in  the same  section we  derive  the constraints
imposed on  the theoretical parameters by  cosmological inflation.  In
Section~\ref{CDM} we  perform a detailed study of  the relic abundance
of the LRHS and  offer numerical estimates of representative scenarios
within the  mSUGRA framework. We also present  numerical estimates for
the scattering cross-section of  the LRHS with the nucleon, indicating
the  presently achieved  and  future sensitivity  of  the current  and
projected  experiments for  DM  searches, such as CDM-II, SuperCDMS and Xenon1T.  Finally,  we summarize  our
conclusions in Section~\ref{conclusions}.

\bigskip

\setcounter{equation}{0}
\section{The {\boldmath $F_D$}-Term Model of Hybrid Inflation}\label{FDmodel}

In this section we first outline the basic structure of the $F_D$-term
model of  hybrid inflation. Then  we briefly review how  the gravitino
abundance  can be  solved  within the  $F_D$-term  model. Finally,  we
present the constraints on the theoretical parameters that are imposed
by CMB data pertinent to  inflation. A more detailed discussion of all
the above issues may be found in~\cite{GPP}.

\subsection{The Model}

The $F_D$-term model may be defined through the superpotential
\begin{equation}
\label{Wmodel}
	W\ =\  \kappa\,\widehat{S}\,
        \Big(\widehat{X}_1\widehat{X}_2\:-\:M^2\Big)\ 
	+\ \lambda\,\widehat{S} \widehat{H}_u  \widehat{H}_d\ 
	+\ \frac{\rho_{ij}}{2}\,\widehat{S}\, \widehat{N}_i\widehat{N}_j\ 
	+\ h^{\nu}_{ij} \widehat{L}_i \widehat{H}_u\widehat{N}_j
 	+\ W_{\rm MSSM}^{(\mu = 0)}\; ,
\end{equation}
where  $\widehat{S}$  is  the  gauge-singlet inflaton  superfield  and
$\widehat{X}_{1,2}$  is  a  chiral  multiplet pair  of  the  so-called
waterfall fields which have  opposite charges under the U(1)$_X$ gauge
group,  i.e.~$Q (\widehat{X}_1)=-Q  (\widehat{X}_2)=1$.   In addition,
$W_{\rm MSSM}^{(\mu  = 0)}$ indicates the  MSSM superpotential without
the $\mu$-term,
\begin{equation} 
	W_{\rm MSSM}^{(\mu=0)}\ =\
	h^u_{ij}\,\widehat{Q}_i\widehat{H}_u\widehat{U}_j\: 
	+\: h^d_{ij}\,\widehat{H}_d\widehat{Q}_i\widehat{D}_j\: 
	+\: h_l\, \widehat{H}_d\widehat{L}_l\widehat{E}_l \; .
\end{equation}
Within the  SUGRA framework, the  sector of soft  supersymmetry (SUSY)
breaking (SSB) derived from~(\ref{Wmodel}) is given by
\begin{equation}\label{Lsoft}
	-\,{\cal L}_{\rm soft} = 
	  M^2_{\tilde S} S^*S
	+ M^2_{\tilde N} N_i^* N_i
	+ \Big(\kappa A_\kappa\, S X_1X_2\: 
	+ \lambda A_\lambda S H_u H_d\: \: 
	+ \frac{\rho}{2}\, A_\rho\, S
	    \widetilde{N}_i\widetilde{N}_i\:
	- \kappa a_S M^2 S \: \ +\ {\rm  H.c.}\,\Big)\,,
\end{equation}
where  $M_{\tilde S}$,  $M_{\tilde N}$,  $A_{\kappa,\lambda,\rho}$ and
$a_S$ are soft SUSY-breaking mass parameters that are all typically of
order  $M_{\rm SUSY}~\sim~1$~TeV.  In addition,  the  $F_D$-term model
contains a FI $D$-term, $-\frac{1}2 g m^2_{\rm FI} D$, associated with
the U(1)$_X$ gauge symmetry of  the waterfall sector. The latter gives
rise to the $D$-term potential
\begin{equation}
  \label{Dterm}
V_D\ =\ \frac{g^2}{8}\; \Big( |X_1|^2\: -\: |X_2|^2\: -\: m^2_{\rm
  FI}\Big)^2\; ,
\end{equation}
where  $g$  is the  U(1)$_X$  gauge-coupling  constant.   The FI  mass
parameter   $m_{\rm  FI}$   is   subdominant  with   respect  to   the
superpotential     tadpole     mass     $M$,    i.e.~$m_{\rm     FI}/M
\stackrel{<}{{}_\sim} 10^{-5}$.

An interesting feature of the $F_D$-term model is the generation of an
effective $\mu$-term  of the required  order $M_{\rm SUSY}$  after the
SSB of  U(1)$_X$. To see  this, let us  neglect the VEVs  of $H_{u,d}$
next to  the large  VEVs of the  waterfall fields  $X_{1,2}$: $\langle
X_{1,2}\rangle = M$. To a good  approximation, the VEV of $S$ may then
be determined by the following part of the potential:
\begin{equation}
  \label{VS}
V_S\ =\ |F_{X_1}|^2\: +\: |F_{X_2}|^2\: +\: M^2_S\,S^*S\: +\:
        \Big[\, \kappa\,M^2 (A_\kappa - a_S)\, S\ +\ {\rm H.c.}\Big]\; ,
\end{equation}
where we have set the waterfall fields $X_{1,2}$ to their actual VEVs.
Substituting the $F$-terms of the waterfall fields,
\begin{equation}
  \label{FX12}
F_{X_{1,2}}\ =\ \kappa S\, \langle X_{2(1)}\rangle\ =\ \kappa M\, S\; ,
\end{equation}
into~(\ref{VS}), we obtain 
\begin{equation}
  \label{VStad}
V_S\ =\ \Big( 2\kappa^2 M^2\: +\: M^2_S \Big)\, S^*S\: +\: 
 \Big[\, \kappa\,M^2 (A_\kappa - a_S)\, S\ +\ {\rm H.c.}\Big]\; .
\end{equation}
It  is then  not difficult  to derive  from~(\ref{VStad}) that  at the
present  epoch of the  Universe, the  inflaton field,  $S$,  acquires the
non-zero VEV
\begin{equation}
  \label{VEVofS}
\langle S \rangle\ =\ 
\frac{1}{2\kappa}\, |A_\kappa - a_S|\: +\: {\cal O}(M^2_{\rm
  SUSY}/M)\; ,
\end{equation}
in  the  phase  convention  that  $\langle  S  \rangle$  is  positive.
Equation~(\ref{VEVofS}) implies the effective $\mu$-term
\begin{equation}
  \label{mu}
	\mu\ =\  
	\lambda\, \langle S  \rangle\ 
	\approx\ \frac{\lambda}{2\kappa}\, |A_\kappa - a_S|\ .
\end{equation}
If  $\lambda \sim  \kappa$, the  size of  $\mu$-parameter is  of order
$M_{\rm  SUSY}$,  as  required  for  a  successful  electroweak  Higgs
mechanism.

In  addition to the  generation of  an effective  $\mu$-parameter, the
third term in~(\ref{Wmodel}), $\frac{1}{2}\, \rho_{ij}\, \widehat{S}\,
\widehat{N}_i   \widehat{N}_j$,    gives   rise   to    an   effective
lepton-number-violating Majorana mass  matrix, i.e.~$M_N = \rho_{ij}\,
v_S$.  If we assume that $\rho_{ij}$ is approximately SO(3) symmetric,
i.e.~$\rho_{ij}  \approx  \rho\;  {\bf  1}_3$, one  obtains  3  nearly
degenerate right-handed neutrinos $N_{1,2,3}$, with mass
\begin{equation}\label{mN}
	m_N\ =\ \rho\, v_S\ .
\end{equation}
If  the  couplings  $\lambda$  and  $\rho$ are  comparable,  then  the
$\mu$-parameter will  set the  scale for the  SO(3)-symmetric Majorana
mass $m_N$, i.e.~$m_N \sim  \mu$~\cite{GP}.  Evidently, this will lead
to a scenario where the singlet neutrinos $N_{1,2,3}$ have TeV or
electroweak-scale masses.   This opens up the  possibility of directly
detecting these singlet Majorana neutrinos through their lepton-number
violating     signatures     at     the     LHC~\cite{NprodLHC}     or
ILC~\cite{NprodILC}.   Furthermore, in  the $F_D$-term  model  the BAU
could   be    explained   by   thermal    electroweak-scale   resonant
leptogenesis~\cite{PU2}.

\subsection{Solution to the Gravitino Overabundance Problem}

The FI mass  term $m_{\rm FI}$ plays a key role  in providing a viable
solution  to the  gravitino  overabundance problem  in the  $F_D$-term
model,  without the  need  to unnaturally  suppress  all the  inflaton
couplings    $\kappa$,     $\lambda$    and    $\rho$     below    the
$10^{-6}$~level~\cite{GP,GPP}.

In detail, the presence of  $m_{\rm FI}$ explicitly breaks an unwanted
discrete symmetry that arises from the permutation of the waterfall fields:
$\widehat{X}_1  \leftrightarrow \widehat{X}_2$.   If $m_{\rm  FI}$ was
absent, the permutation symmetry would remain exact even after the SSB
of the U(1)$_X$.  This would  act like parity and was therefore termed
$D$-parity in~\cite{GP}.  As a consequence of $D$-parity conservation,
the $D$-odd waterfall particles of  mass $g M$ would have been stable,
and     if    abundantly     produced     during    the     preheating
epoch~\cite{PREHEATING,GBRM},  they could  overclose  the Universe  at
late times.

To avoid this undesirable situation, we introduce a small but non-zero
FI term  $m_{\rm FI}$. In  this case, the $D$-odd  waterfall particles
will have forbidden decays  to two $D$-even inflaton-related fields of
mass $\kappa  M$, induced by the  FI term. To  kinematically allow for
such decays, we assume that $\kappa  < g/2$, where $g$ is the
value of  the U(1)$_X$ coupling constant  at the GUT  scale.  The late
decays  of the  $D$-odd waterfall  fields will  then reheat  again the
Universe at  temperature $T_g$, and  so release enormous  entropy that
might    be   sufficient   to    reduce   the    gravitino   abundance
$Y_{\widetilde{G}}$ below the BBN  limits.  More explicitly, after the
Universe  passes  through  a  second reheating  phase,  the  gravitino
abundance may be estimated by~\cite{GPP}:
\begin{equation}
  \label{YGtilde}
Y_{\widetilde{G}}\ \approx \ \frac{7.6\times 10^{-11}}{\kappa g}\
\Bigg( \frac{T_g}{10^{10}~{\rm GeV}}\Bigg)\; ,
\end{equation}
Hence, for  second reheat temperatures  $T_g \sim 1$~TeV  and inflaton
couplings $\kappa  \stackrel{>}{{}_\sim} 10^{-2}$, the strict
constraint $Y_{\widetilde{G}} \stackrel{<}{{}_\sim} 10^{-15}$, for \(m_{\tilde G}\lesssim 500\)~GeV, can be
comfortably met.

To  determine the  second reheat  temperature  $T_g$, we  may use  the
standard freeze-out condition $\Gamma_g = H(T_g)$, where
\begin{equation}
  \label{Gg}
\Gamma_g\ =\ \frac{g^4}{128\pi}\ \frac{m^4_{\rm FI}}{M^3}
\end{equation}
is the  decay rate of the  $D$-odd particles and 
\begin{equation}
  \label{Hrad}
H(T)\ =\ \Bigg( \frac{\pi^2 g_*}{90}\Bigg)^{1/2}\, \frac{T^2}{m_{\rm Pl}}
\end{equation}
is the  Hubble expansion parameter  in the radiation dominated  era of
the Universe and  $m_{\rm Pl} = 2.4\times 10^{18}$~GeV  is the reduced
Planck mass. In  particular, for a fixed given value  of~$T_g$, we may
infer the required size of the FI mass term $m_{\rm FI}$~\cite{GPP}:
\begin{equation}
  \label{mFI}
\frac{m_{\rm FI}}{M}\ \approx\ 8.4 \times 10^{-4}\times \Bigg(
\frac{0.5}{g}\Bigg)^{3/4} \Bigg(\frac{T_g}{10^9~{\rm
GeV}}\Bigg)^{1/2}\, \Bigg( \frac{10^{16}~{\rm
GeV}}{M}\Bigg)^{1/4}\; .
\end{equation}
As can  be seen from~(\ref{mFI}), for  $T_g \sim 1$~TeV,  it should be
$m_{\rm FI}/M \sim 10^{-6}$, so the FI mass term $m_{\rm FI}$ needs be
much smaller  than $M$. Detailed  discussion of how such  an hierarchy
can  be naturally  achieved within  the SUGRA  framework may  be found
in~\cite{GPP}.

\subsection{Constraints from Cosmological Inflation}

Here we recall the constraints derived in~\cite{GPP} on the $F_D$-term
model  from   cosmological  inflation.   In  fact,  there   are  three
constraints that need to be considered.

The  first  constraint arises  from  the  requirement  of solving  the
horizon  and flatness  problems  of the  standard Big-Bang  Cosmology.
According to  the inflationary paradigm, these  problems may naturally
be solved, if our observable  Universe had an accelerated expansion of
a  number of 50--60  $e$-folds.  In  the slow-roll  approximation, the
number of $e$-folds, ${\cal N}_e$, may be calculated by~\cite{review}
\begin{equation}
  \label{Nefold}
{\cal N}_e\ =\ \frac{1}{m^2_{\rm Pl}}\; \int_{\phi_{\rm
    end}}^{\phi_{\rm exit}}\, d\phi\: \frac{V_{\rm inf}}{V'_{\rm inf}}\
    \simeq\ 55\; ,
\end{equation}
where $\phi=\sqrt{2}\, {\rm Re}\,S$  is the inflaton field and $V_{\rm inf}$
is the $F_D$-term inflaton potential  that can be found in Section 2.1
of~\cite{GPP}. We  will always denote differentiation  with respect to
$\phi$ with a prime on  $V_{\rm inf}$.  Moreover, $\phi_{\rm exit}$ is
the value of $\phi$, when our present horizon scale exited inflation's
horizon, whilst $\phi_{\rm end}$ is its value at the end of inflation.
Specifically, the field value  $\phi_{\rm end}$ may be determined from
the condition:
\begin{equation}
  \label{slow}
{\sf max}\{\epsilon(\phi_{\rm end}),|\eta(\phi_{\rm
end})|\}\ =\ 1\, ,
\end{equation}
with
\begin{equation}
  \label{epseta}
\epsilon\ =\ \frac{m_{\rm Pl}^2}{2}\ \left(
\frac{V'_{\rm inf}}{V_{\rm inf}}\right)^2\,,\qquad
\eta\ =\  m_{\rm Pl}^2\ \frac{V''_{\rm inf}}{V_{\rm inf}}\ .
\end{equation}

The other  two inflationary constraints come from  the so-called power
spectrum  $P_{\cal R}$  of  curvature perturbations  and the  spectral
index $n_s$.   The square root  of the power  spectrum, $P^{1/2}_{\cal
R}$, is given by
\begin{equation}
    \label{PR}
P^{1/2}_{\cal R}\ =\ \frac{1}{2\sqrt{3}\, \pi m^3_{\rm Pl}}\;
\frac{V_{\rm inf}^{3/2}(\phi_{\rm exit})}{|V'_{\rm inf}(\phi_{\rm
exit})|}\ .
\end{equation}
This prediction must be compared  with the result obtained by a 3-years
WMAP analysis of CMB data~\cite{WMAP3},
\begin{equation}
  \label{Pr}
P^{1/2}_{\cal R}\ \simeq\ 4.86\times 10^{-5}\, .
\end{equation}
Moreover, in  the slow-roll approximation, the  spectral index $n_{\rm
s}$ is given by~\cite{review}
\begin{equation}
  \label{nS}
n_{\rm s}\ =\ 1-6\epsilon(\phi_{\rm exit})\ +\ 2\eta(\phi_{\rm exit})\
\simeq\ 1\ +\ 2\eta(\phi_{\rm exit}),
\end{equation}
where the parameter $\epsilon$  is negligible in the $F_D$-term model.
Recently, after analysing its data collected in the last 5 years, WMAP
has reported the value for the spectral index~\cite{Dunkley:2008ie}:
\begin{equation}
  \label{nswmap}
n_{\rm  s}\: -\: 1\ =\ -0.037_{-0.015}^{+0.014}\ .
\end{equation}
This result  slightly favours a red-tilted spectrum  and is consistent
with scale invariance at the 2.64~$\sigma$ confidence level.

Given   the   three   constraints   (\ref{Nefold}),   (\ref{Pr})   and
(\ref{nswmap}), and  assuming that  all inflaton couplings  are equal,
i.e.~$\kappa =  \lambda = \rho$,  one obtains within mSUGRA  the upper
bound~\cite{GPP}
\begin{equation}
  \label{mSUGRA}
\kappa\ \stackrel{<}{{}_\sim}\ 2\times 10^{-2}\; .
\end{equation}
On the  other hand,  the inflationary  scale $M$ is  close to  the GUT
scale,  i.e.~$M \sim  10^{16}$~GeV,  when $\kappa$  reaches its  upper
bound imposed  by inflation.  For  an inflaton sector that  realizes a
next-to-minimal K\"ahler potential with a negative Hubble-induced mass
term  for  $S$~\cite{hilltop}, the  upper  limit  on  $\kappa$ may  be
slightly relaxed to~\cite{GPP}
\begin{equation}
  \label{nmSUGRA}
\kappa\ \stackrel{<}{{}_\sim}\ 3.2\times 10^{-2}\; ,
\end{equation}
whilst $M$ decreases to $M \simeq 0.5 \times 10^{16}$~GeV.

It is important to  properly translate the upper bounds~(\ref{mSUGRA})
and~(\ref{nmSUGRA}) on $\kappa$ obtained at the inflationary scale $M$
into  the  respective  ones  on  $\lambda$ and  $\rho$  for  the  soft
SUSY-breaking scale~$M_{\rm SUSY}$. As  we will see more explicitly in
the next  section, it  is the product  $\lambda\rho$ evaluated  at the
scale $M_{\rm SUSY}$ that controls the strength of annihilation of the
LRHSs into the Higgs fields  and other SM particles.  Even though the
renormalization  group (RG) evolution  of $\rho$  from $M$  to $M_{\rm
SUSY}$ may be ignored, as~$\rho (M) \approx \rho (M_{\rm SUSY})$, this
is  not the  case for  the coupling  $\lambda$.  Neglecting  gauge and
small  Yukawa  couplings  of  order  $10^{-1}$, the  RG  equation  for
$\lambda$ is given by~\cite{BS}
\begin{equation}
  \label{RGlambda}
16\pi^2\, \frac{d\lambda}{dt}\ =\ \lambda\, \Bigg(\, \frac{3}{2}\,
h^2_t\: +\: \frac{3}{2}\: h^2_b\,\Bigg)\; ,
\end{equation}
where $t = \ln (Q^2/M_{\rm  SUSY})$. Assuming that the RG evolution is
dominated  by the  top-quark Yukawa  coupling $h_t$,  the  solution to
(\ref{RGlambda}) is easily found to be
\begin{equation}
  \label{RGsolution}
\lambda (M_{\rm SUSY})\ =\ \lambda (M)\, 
\Bigg(\frac{M_{\rm SUSY}}{M}\Bigg)^{3h^2_t/(16\pi^2)}\ \approx \
0.57\times \lambda (M)\; .
\end{equation}
To obtain the last  result in~(\ref{RGsolution}), we assumed that $h_t
\approx 1$ and $M_{\rm SUSY}/M \sim 10^{-13}$. Then, starting with the
boundary condition  $\lambda = \kappa$ at the  inflationary scale $M$,
the  RG  running~(\ref{RGsolution})  of  $\lambda$ implies  the  upper
limits:
\begin{equation}
  \label{Ulambda}
\lambda (M_{\rm SUSY})\ \stackrel{<}{{}_\sim}\ 1.14\times
  10^{-2}\,,\qquad \lambda (M_{\rm SUSY})\ \stackrel{<}{{}_\sim}\ 
1.82\times 10^{-2}\; ,
\end{equation}
for an inflaton  sector with a minimal and  a next-to-minimal K\"ahler
potential, respectively.

In  addition   to  constraints  from  inflation,  one   may  also  get
constraints on the  size of $M$ from cosmic strings  that arise due to
the SSB  of the local U(1)$_X$  symmetry.  For values  of $\kappa \sim
10^{-2}$ of  our interest, this implies that  one must have~\cite{GPP}
$M \stackrel{<}{{}_\sim} 0.5\times  10^{16}$~GeV.  This constraint may
be a  bit restrictive for the  mSUGRA model, but it  can be completely
avoided if  the waterfall sector realizes an  SU(2)$_X$ gauge symmetry
instead   of   U(1)$_X$,    whose   SSB   generates   no   topological
defects~\cite{GPP}.  Consequently, we will conservatively consider the
limits  stated in~(\ref{mSUGRA}),  (\ref{nmSUGRA}) and~(\ref{Ulambda})
when implementing  inflationary constraints on the  relic abundance of
the LRHS in the next section.

\bigskip

\setcounter{equation}{0}
\section{Right-Handed Sneutrino as Thermal Dark Matter}\label{CDM}

In the  $F_D$-term hybrid model  $R$-parity is conserved,  even though
the lepton number $L$, as well  as $B-L$, are explicitly broken by the
Majorana   operator    $\frac{1}{2}   \rho   \widehat{S}   \widehat{N}
\widehat{N}$. We note that all  superpotential couplings either  conserve the $B-L$
number or break it by even  number of units.  Since $R$-parity of each
superpotential operator is determined to  be $R = (-1)^{3(B-L)} = +1$,
the $F_D$-term  hybrid model conserves $R$-parity.   As a consequence,
the LSP  of the spectrum is stable  and can be a  viable candidate for
Cold Dark  Matter~(CDM). As  an extension of  the MSSM, our  model can
accommodate  the standard  SUSY CDM  candidates, such  as  the lightest
neutralino. Because  of the connection between the  Higgs and neutrino
sectors, on  the one  hand, and  inflation, on the  other, it  is very
interesting  to  explore  the  possibility of  having  a  right-handed
sneutrino as LSP in order to solve the CDM problem.  As we will see in
Section 3.3, this renders  the $F_D$-term model much more constrained,
leading to sharp predictions for scattering cross-sections relevant to
experiments of direct searches for~CDM.

\subsection{Sneutrino Mass Spectrum}

Before  calculating the  sneutrino relic  abundance in  our  model, we
first observe that light  right-handed sneutrinos may easily appear in
the spectrum.  Ignoring the terms proportional to the small neutrino-Yukawa  couplings, the
\(6\times    6\)   right-handed    sneutrino   mass    matrix   ${\cal
M}^2_{\widetilde  N}$ is  given in  the weak  basis $(\widetilde{N}_{1,2,3},
\widetilde{N}^*_{1,2,3})$ by
\begin{equation}
  \label{eq:SneutrinoMassMatrix}
	{\cal M}^2_{\widetilde N} = 
	\frac{1}{2} 
	\left(\begin{array}{cc}
		\rho^2 v^2_S      + M^2_{\widetilde  N} & 
		\rho A_\rho v_S   + \rho\lambda v_u v_d \\
		\rho A^*_\rho v_S + \rho\lambda v_u v_d &
		\rho^2 v^2_S      + M^2_{\widetilde  N}
	\end{array} \right),
\end{equation}
where  $v_S   =  \langle  S  \rangle$,  $v_{u,d}   =  \langle  H_{u,d}
\rangle$. Moreover, $M^2_{\widetilde  N}$ is the soft  SUSY-breaking mass matrix
associated with  the sneutrino fields and \(A_\rho\)  is the sneutrino
trilinear coupling  matrix. In general,  \({\cal M}^2_{\widetilde N}\)
is diagonalized by a unitary matrix \(U_{\widetilde N}\) such that
\begin{equation}\label{eq:SneutrinoDiag}
	U_{\widetilde N}^\dagger  \,
	{\cal M}^2_{\widetilde N} \,
	U_{\widetilde N} =
	\textrm{diag}
	\left(m_{\tilde N_1}^2,m_{\tilde N_2}^2,\dots,m_{\tilde N_6}^2\right),
\end{equation}
where the sneutrino masses are ordered, such that \(m_{\tilde N_1} < m_{\tilde N_2}< \dots < m_{\tilde N_6}\). Neglecting the possible
flavor   structure   contained   in   the   \(3\times   3\)   matrices
\(M^2_{\widetilde  N}\) and  \(A_\rho\), the  sneutrino  spectrum will
then consist of 3 light (heavy) right-handed sneutrinos with masses
\begin{equation}\label{eq:SneutrinoMasses}
	m^2_{\tilde N_{L(H)}} = 
	     \rho^2 v^2_S 
	+    M^2_{\widetilde N}  
	-(+) \left|\rho A_\rho v_S + \rho\lambda v_u v_d\right|.
\end{equation}
All    mass   terms    in   (\ref{eq:SneutrinoMasses})    are   ${\cal
O}(100$--1000)~GeV,  so  a  proper  choice  of  model  parameters  can
accommodate  a LRHS  to act  as  LSP.  Unless  the trilinear  coupling
\(A_\rho\) is small compared  to \(\mu\), the off-diagonal elements in
(\ref{eq:SneutrinoMassMatrix}) will  induce a sizeable  mixing between
the  heavy and  light right-handed  sneutrino states,  suppressing the
light masses to  values smaller than \((\mu^2+M_{\tilde N}^2)^{1/2}\).
This will be  demonstrated in our discussion of  the numerical results
in Section~3.3,  where the  $F_D$-term model is  embedded within the mSUGRA
framework.

\subsection{Sneutrino Annihilation and Relic Density}

Right-handed sneutrinos  as CDM  were considered in~\cite{GGP}  in the
context   of   the  MSSM   with   right-handed  neutrino   superfields
$\widehat{N}_i$  and  bare  Majorana  masses  $M^N_{ij}  \widehat{N}_i
\widehat{N}_j$.   This   analysis  shows  that   thermal  right-handed
sneutrinos  have  rather  high  relic abundances  and  will  generally
overclose  the Universe.   The reason  is  that because  of the  small
neutrino Yukawa couplings $h^\nu_{ij}$,  the self- and co-annihilation
interactions of the sneutrino LSP with itself and other MSSM particles
are rather weak.  These weak  processes do not allow the sneutrino LSP
to  stay long  enough  in thermal  equilibrium  before its  freeze-out
temperature, such that its number density gets reduced to the observed
value   $\Omega_{\rm  DM}   h^2   \approx  0.11$~\cite{Dunkley:2008ie}
[cf. (\ref{OmegaDM})].  In fact, the predicted values for $\Omega_{\rm
DM}  h^2$ turn  out  to be  many  orders of  magnitude larger  than~1.
Instead, right-handed  sneutrinos can be viable  thermal DM candidates in the MSSM if
they  significantly mix  with  left-handed sneutrinos,  either by increasing the 
SUSY-breaking  trilinear couplings~\cite{Hooper:2004dc}~\footnote{For an earlier discussion, see also
the paper by N.   Arkani-Hamed {\em et al.}  in~\cite{Francesca}.}, or by lowering the right-handed neutrino mass scale~\cite{Arina:2008bb}. 
Alternatively, right-handed sneutrinos may become thermal DM by introducing a new U(1)' gauge coupling to make the self-annihilation interaction sufficiently strong~\cite{LMN}. Recently, there has been a paper~\cite{Cerdeno} discussing the possibility of right-handed sneutrinos as DM in an extended version of the next-to-minimal supersymmetric standard model.

\begin{figure}[t]
\centering
\includegraphics[clip,width=0.90\textwidth]{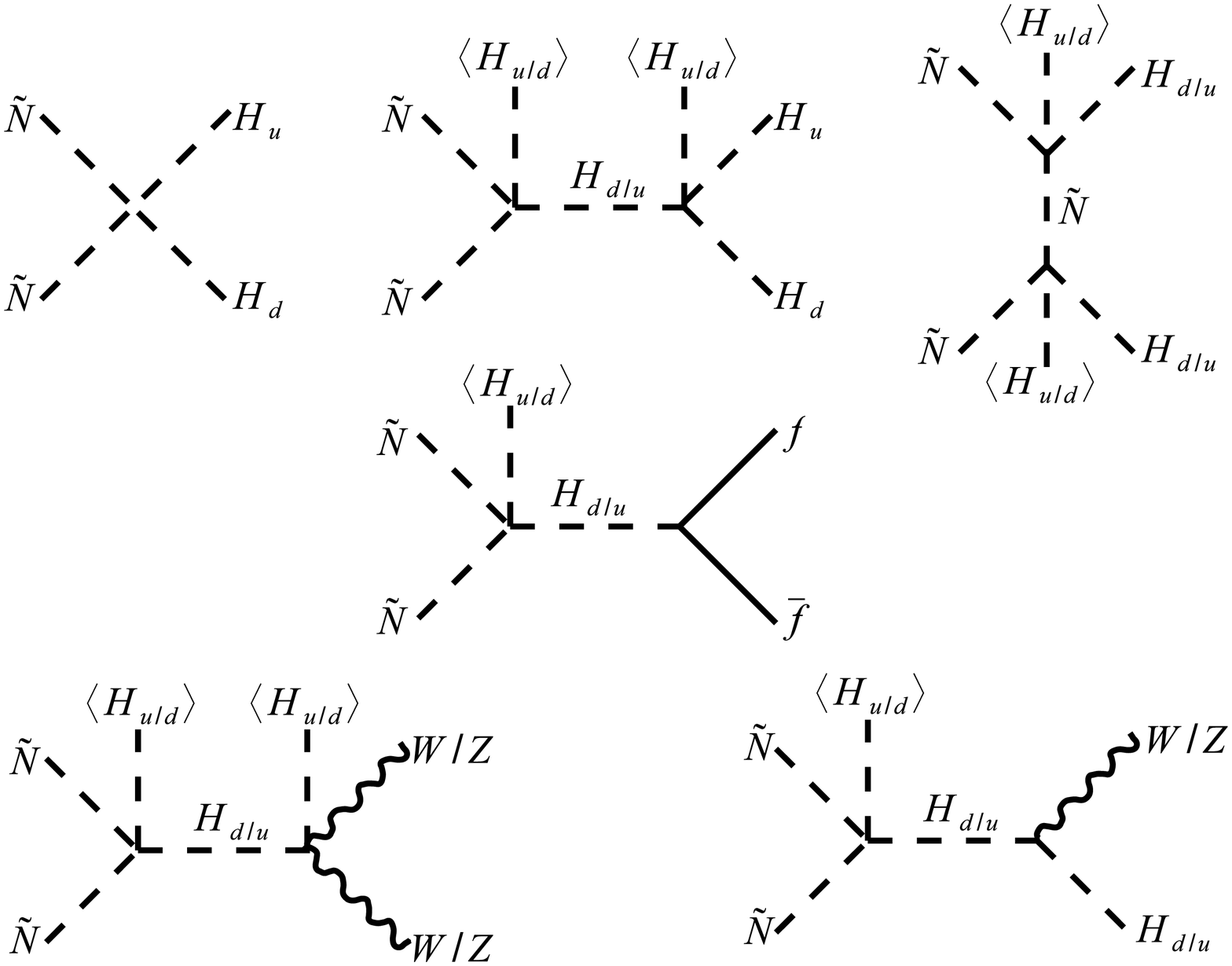}
\caption{Feynman graphs related to sneutrino annihilation.}
\label{fig:graphs}
\end{figure}

In the $F_D$-term hybrid model, a novel possibility opens up.
As was first observed in~\cite{GPP}, there   exists    a   new   quartic   coupling    described   by   the
Lagrangian~\footnote{The    implications   of    a   generic
singlet-Higgs  quartic coupling  for the  CDM abundance  and detection
were studied  before in~\cite{Silveira:1985rk,mcdonald,pospelov}, within a simple
non-SUSY model.}
\begin{equation}\label{Llsp}
	{\cal L}^{\rm LSP}_{\rm int} =
	\frac{1}{2}\lambda\rho \tilde N^*_i \tilde N^*_i H_uH_d 
	+{\rm H.c.}\; .
\end{equation}
This quartic coupling between right-handed sneutrinos and Higgs fields
results from  the $F$-term of  the inflaton field  $F_S$: $\frac{1}{2}
\rho \widehat{N}_i \widehat{N}_i + \lambda \widehat{H}_u \widehat{H}_d
\subset  F_S$.   If strong  enough,  the interaction~(\ref{Llsp})  can
thermalize  the  sneutrinos  and  make  them  annihilate  to  a  level
compatible  with the current  CMB  data  via the  processes  depicted  in
Figure~\ref{fig:graphs}.

For sneutrino masses of our  interest, the most relevant processes are
the  off-resonant  pair-production  of  $W$ bosons  and  the  on-shell
pair-production  of light  Higgs bosons.  An initial  estimate  of the
process  \(\tilde N\tilde N  \to \langle  H_u\rangle H_d  \to W^+W^-\)
for \(m_{\tilde N} > m_W\) yields
\begin{equation}
  \label{estimate1}
	\Omega_{\rm DM}\, h^2\ \approx\
	\left(\frac{10^{-4}}{\rho^2\lambda^2}\right)
	\left(\frac{\tan\beta\, m_H}{g_W\, m_W}\right)^2\; .
\end{equation}
In  order  to  obtain  an  acceptable CDM  density,  relatively  large
couplings \(\rho\) and  \(\lambda\) are needed, \(\rho\lambda \gtrsim
0.1\).  However, these  large values for $\lambda$ and  $\rho$ are not
compatible with  the constraints  derived by inflation.  

The situation  differs for sneutrino masses $m_{\widetilde  N} < m_W$,
in  large $\tan\beta$ scenarios,  in which  light Higgs  bosons couple
appreciably  to  $b$-quarks~\cite{Sabine}.    In  particular,  in  the
kinematic region $m_{H_1}  \approx 2 m_{\widetilde{N}_{1}}$, the
self-annihilation process  $\widetilde{N}_{1} \widetilde{N}_{1} \to \langle  H_u \rangle H_d \to b\bar{b}$  becomes resonant, and
the above estimate modifies to
\begin{equation}
  \label{estimate2}
	\Omega_{\rm DM} h^2 \approx 10^{-4} \times B^{-1}(H_1 \to
	\widetilde N_1 \widetilde N_1) \times
	\left(\frac{m_{H_1}}{100~{\rm GeV}}\right)^2.
\end{equation}
Consequently,  if  the couplings  $\lambda,\rho$  are  not too  small,
e.g.~$\lambda\rho  \gtrsim  10^{-3}$,  the  right-handed  sneutrino
$\widetilde{N}_1$  can   now  efficiently  annihilate   via  a  resonant
$H_1$-boson  into  pairs of  $b$-quarks,  thus  obtaining  a relic  DM
density compatible with the observed value (\ref{OmegaDM}).

We   will  now   show  that   the   naive  estimates~(\ref{estimate1})
and~(\ref{estimate2}) presented in~\cite{GPP}  are in a fairly  good 
agreement  with  a complete  calculation  of  all relevant  sneutrino
annihilation processes displayed  in Figure~\ref{fig:graphs}.  To this
end,  we use  the short-hand  notation \(M_{XY}  =  M(\tilde N_a\tilde
N_b\to  XY)\)  to  denote  the  individual  matrix  elements  for  the
annihilation  of sneutrinos  \(\tilde N_a\)  and \(\tilde  N_b\).  The
contributing   processes    may   be   listed    as   follows   (\(c_w
=\cos\theta_w\), \(v=2m_W/g_w\)):
\begin{itemize}
\item[{\bf (i)}]  \(\tilde N_a  \tilde N_b \longrightarrow  H^+ H^-\),
  via contact quartic interaction and $s$-channel Higgs exchange:
\begin{equation}
	M_{H^+H^-}\ =\
	g_{\tilde N_a \tilde N_bH^+H^-}
	- v^2 \sum_{k=1}^3\:
	\frac{g_{\tilde N_a \tilde N_b H_k}g_{H_k H^+H^-}}
	{s-m_{H_k}^2+im_{H_k}\Gamma_{H_k}}\ ; 
\end{equation}

\item[{\bf (ii)}] \(\tilde N_a \tilde N_b \longrightarrow W^+ W^-\), via
$s$-channel Higgs exchange:
\begin{equation}
	M_{W^+W^-}\ =\
	g_w m_W v\, 
	\Bigg[\, 2\: +\: \Bigg(1-\frac{s}{2m_W^2}\Bigg)^2\,\Bigg]^{1/2}\:
	\sum_{k=1}^3\:
	\frac{g_{\tilde N_a \tilde N_b H_k}g_{H_k VV}}
	{s-m_{H_k}^2 + i m_{H_k}\Gamma_{H_k}}\ ;
\end{equation}

\item[{\bf (iii)}] \(\tilde N_a \tilde N_b \longrightarrow ZZ\), via
$s$-channel Higgs exchange:
\begin{equation}
	M_{ZZ}\ =\
	\frac{g_w m_W v}{2c_w}\, 
	\Bigg[\, 2\: +\: \Bigg(1-\frac{s}{2m_Z^2}\Bigg)^2\,\Bigg]^{1/2}\:
	\sum_{k=1}^3
	\frac{g_{\tilde N_a \tilde N_b H_k} g_{H_k VV}}
	{s - m_{H_k}^2 + im_{H_k}\Gamma_{H_k}}\ ;
\end{equation}

\item[{\bf (iv)}] \(\tilde N_a \tilde N_b \longrightarrow f \bar f\), via
$s$-channel Higgs exchange:
\begin{equation}
	M_{f_\alpha \bar f_\alpha}\ =\
	v\sqrt{2s}\,
	\Bigg[\,
		|A_S|^2\Bigg(1-\frac{4m_\alpha^2}{s}\Bigg)\: +\: |A_P|^2\,
	\Bigg]^{1/2}\ ,
\end{equation}
with
\begin{equation}
	A_{S/P}\ =\ 
	\sum_{k=1}^3\:  
   \frac{
		g_{\tilde N_a \tilde N_b H_k}
		g_{f_\alpha}g^{S/P}_{H_k \bar f_\alpha f_\alpha}
	}
	{s-m_{H_k}^2+im_{H_k}\Gamma_{H_k}}\ ;
\end{equation}

\item[{\bf (v)}] \(\tilde N_a \tilde N_b \longrightarrow H_i H_j\), via
  contact quartic interaction, $s$-channel Higgs exchange and
  $t/u$-channel sneutrino exchange:
\begin{eqnarray}
	M_{H_i H_j} \!&=&\!
	g_{N_aN_bH_iH_j}\: 
	-\: v^2\,
	\sum_{k=1}^3\: 
	\frac{g_{\tilde N_a \tilde N_b H_k}g_{H_i H_j H_k}}
	{s-m_{H_k}^2+im_{H_k}\Gamma_{H_k}} \nonumber\\
	&-& v^2\,
	\sum_{c=1}^6\: 
        \frac{g_{\tilde N_a \tilde N_c H_i}
					g_{\tilde N_b \tilde N_c H_j}}
	{t-m_{\tilde{N}_c}^2}\ 
	-\ v^2\,
	\sum_{c=1}^6
	\frac{g_{\tilde N_a \tilde N_c H_j}
			g_{\tilde N_b \tilde N_c H_i}}
	{u-m_{\tilde{N}_c}^2}\ ;
\end{eqnarray}

\item[{\bf (vi)}] \(\tilde N_a \tilde N_b \longrightarrow H^+ W^-\),
via $s$-channel Higgs exchange:
\begin{equation}
	M_{H^+W^-}\ =\
	\frac{g_w v}{2}\,
	\Bigg[\,
		\frac{s^2}{4m_W^2}
		\Bigg(1-\frac{m_W^2+m_{H^+}^2}{s}\Bigg)^2\: -\: m_{H^+}^2
	\Bigg]^{1/2}
	\sum_{k=1}^3
	\frac{g_{\tilde N_a \tilde N_b H_k}
			g_{H_k H^+ W^-}}
	{s-m_{H_k}^2+im_{H_k}\Gamma_{H_k}}\ ;
\end{equation}

\item[{\bf (vii)}] \(\tilde N_a \tilde N_b \longrightarrow H_i Z\),
via $s$-channel Higgs exchange:
\begin{equation}
	M_{H_i Z}\ =\
	\frac{g_w v}{4c_s}\,
	\Bigg[\,
		\frac{s^2}{4m_Z^2}
		\Bigg(1-\frac{m_Z^2+m_{H_i}^2}{s}\Bigg)^2\: -\: m_{H_i}^2
	\Bigg]^{1/2}
	\sum_{k=1}^3
	\frac{g_{\tilde N_a \tilde N_b H_k}g_{H_k H_i Z}}
	{s-m_{H_k}^2+im_{H_k}\Gamma_{H_k}}\ .
\end{equation}
\end{itemize}
In the  above, the effective  sneutrino-to-Higgs couplings \(g_{\tilde
N_a \tilde  N_b H^+H^-}\), \(g_{\tilde  N_a \tilde N_b H_j  H_j}\) and
\(g_{\tilde  N_a \tilde  N_c H_i}\)  that arise  from  the interaction
Lagrangian~(\ref{Llsp})    are    given    by    (\(c_\beta=\cos\beta,
s_\beta=\sin\beta\))
\begin{eqnarray}
	g_{\tilde N_a \tilde N_b H^+H^-} \!&=&\! 
	\frac{\lambda\rho}{2}\: c_\beta s_\beta\delta_{ab}, \\
	g_{\tilde N_a \tilde N_b H_i H_j} &=&
	\frac{\lambda\rho}{2}\:  
	\frac{\delta_{ab}}{1+\delta_{ij}}
	\left[\left(
		  O_{\phi_u i} O_{\phi_d j} 
		+ O_{a i}      O_{\phi_u j}s_\beta
		+ O_{a i}      O_{\phi_d j}c_\beta
		- O_{a i}      O_{a j}s_\beta c_\beta
	\right)\right. \nonumber\\
	&&\qquad\qquad\quad+ \left.(i \leftrightarrow j)
	\right], \\
	g_{\tilde N_a \tilde N_b H_i}  \!&=&\! 
	\frac{\lambda\rho}{2}\:
	\left(
		O_{\phi_d i}s_\beta + O_{\phi_u i}c_\beta
	\right)\delta_{ab}\; ,
\end{eqnarray}
where $O$ is  the $3\times 3$ Higgs-boson mixing  matrix, defined such
that 
\begin{equation}
(\phi_d,\, \phi_u,\, a)^T\ =\ O \: (H_1,\, H_2,\, H_3)^T\; .
\end{equation}
For the  effective Higgs-boson  couplings \(g_{H_k H_i  Z}\), \(g_{H_k
H^+ W^-}\),  \(g_{H_i H_j  H_k}\), \(g_{H_k H^+H^-}\),  \(g_{H_k VV}\)
and   \(g_{f_\alpha}\,  g^{S/P}_{H_k   f_\alpha  \bar   f_\alpha  }\),
including $O$, the Higgs-boson  masses $m_{H_{1,2,3}}$ and their decay
widths $\Gamma_{H_{1,2,3}}$,  we follow the  notations and conventions
of~\cite{Lee:2003nta,Lee:2007gn}  and calculate them  by means  of the
computational   package  {\tt CPsuperH}.   

The   total  annihilation   cross-section  \(\sigma_{ab}=\sigma(\tilde
N_a\tilde N_b\to \textrm{all})\) may then be conveniently expressed as
the sum of all channels,
\begin{equation}
	\sigma_{ab}\ =\
	\sigma_{H^+H^-}+
	\sigma_{W^+W^-}+
	\sigma_{ZZ}+
	\sigma_{H^+W^-}+\sigma_{H^-W^+}+
	\sum_{i=1}^3        \sigma_{H_i Z} +
	\sum_{i,j=1}^3     \sigma_{H_i H_j} + 
	\sum_{f=\tau,b,t} \sigma_{f \bar f}\; .
\end{equation}
The individual cross sections \(\sigma_{XY}\) are defined by
\begin{equation}
	\sigma_{XY} \ = \ 
	\frac{1}{1+\delta_{XY}}\:
	\frac{1}{16\pi \lambda(s,m_{\tilde N_a}^2,m_{\tilde N_b}^2)}\:
	\int_{t^-}^{t^+} dt\, |M_{XY}|^2,
\end{equation}
with
\begin{eqnarray}
	t^\pm \!&=&\!  
	m_X^2 + m_{\tilde N_a}^2 - 
	\frac{1}{2s}
	\left(
		(s+m_{\tilde N_a}^2-m_{\tilde N_b}^2)(s+m_X^2-m_Y^2)
	\right. \nonumber\\
		&& \qquad\qquad\qquad\qquad
	\left.\mp \lambda^{1/2}(s,m_{\tilde N_a}^2,m_{\tilde N_b}^2)
		    \lambda^{1/2}(s,m_X^2,m_Y^2)\right)\;,\qquad\quad \\
	\lambda(a,b,c) \!&=&\! (a-b-c)^2-4bc.
\end{eqnarray}
In     order     to     calculate     the    relic     density,     we
follow~\cite{Griest:1990kh}   and  use   an   effective  cross-section
averaged over all initial sneutrino channels,
\begin{eqnarray}
  \label{eq:EffectiveCrossSection}
	\sigma_{\rm eff} \! &=&\!
	\sum_{a,b=1}^6	\sigma_{ab} \frac{g_a g_b}{g_{\rm eff}^2}
		(1+\Delta_a)^{3/2}
		(1+\Delta_b)^{3/2}\;
\exp \big[ -x(\Delta_a+\Delta_b)\big]\; ,
\end{eqnarray}
where
\begin{equation}
g_{\rm eff} \ = \
 \sum_{a=1}^6 g_a(1+\Delta_a)^{3/2}e^{-x\Delta_a}\; , \qquad
\Delta_a \ =\ \frac{m_{\tilde N_a}-m_{\tilde N_1}}{m_{\tilde N_1}}\ .
\end{equation}
In~(\ref{eq:EffectiveCrossSection}),   both   the   effects   of   LSP
self-annihilation and co-annihilation  with the heavier sneutrinos are
included\footnote{Note    that    co-annihilation    effects    become
significant, only if the  mass differences with the heavier sneutrinos
are smaller or comparable to the LSP freeze-out temperature, i.e,~when
\(m_{\tilde  N_a}-m_{\tilde   N_b}\stackrel{<}{{}_\sim}  T_f\).}.   In
terms of the effective cross-section~(\ref{eq:EffectiveCrossSection}),
the thermally-averaged effective cross-section may be calculated as
\begin{equation}\label{eq:sigmavEffective}
	\langle\sigma v\rangle\ =\ \frac{x^{3/2}}{2\pi^{3/2}}
	\int_0^\infty dv\, v^2 (\sigma_{\rm eff} v)\, e^{-x v^2/4}\ ,
\end{equation}
where the integrand is expressed in terms of the relative velocity \(v\), such that
\begin{equation}
	s \ = \ \frac{4m_{\tilde N_1}^2}{1-v^2/4} \ . 
\end{equation}
From the expression (\ref{eq:sigmavEffective}), we may determine the freeze-out temperature
 \(x_f= m_{\tilde N_1}/T_f\) by iteratively solving the equation
\begin{equation}
	x_f\ =\ 
	\ln\Bigg(
\frac{0.038 g_{\rm eff}\, M_{\rm Pl}\, m_{\tilde N_1} \langle\sigma v\rangle}
	{g_*^{1/2}x_f^{1/2}}\Bigg)\; , 
\end{equation}
where  \(  M_{Pl}=1.22\times  10^{19}\)~GeV  is the  Planck  mass  and
\(g_*\)  is the  total  number of  effective  relativistic degrees  of
freedom at  the temperature  of the LSP  freeze-out.  The  present day
relic density is then given by
\begin{equation}\label{eq:RelicDensity}
	\Omega_{\rm DM}\, h^2\ \approx\ 
	\frac{1.07\times 10^9~\textrm{GeV}^{-1}}{J\, g_*^{1/2} m_{Pl}}
\end{equation}
where $J$ is the  post freeze-out annihilation efficiency factor given
by
\begin{equation}
	J\ =\ \int_0^\infty dv\, v (\sigma_{eff}v)\, 
         \textrm{erfc}(v\sqrt{x_f}/2).
\end{equation}
In our  numerical estimates,  we neglect the  flavor structure  of the
right-handed  sneutrinos  and   treat  the  three  light  right-handed
sneutrinos \(\tilde N_{1,2,3}\) as being  essentially degenerate~\footnote{Note that the second and third right-handed sneutrinos \(\tilde N_{2,3}\) will decay to the LRHS \(\tilde N_1\) through the processes~\(\tilde N_{2,3}\to\tilde N_1 \gamma, \tilde N_1 \nu\bar\nu\). We do not address potential problems for BBN from the late decays of~\(\tilde N_{2,3}\), since their rates strongly depend on the flavor structure of \(\rho_{ij}\) and the Yukawa couplings \(h^\nu_{ij}\) [cf. (\ref{Wmodel})] and on the details of the model in general.}.   Since all  three light
sneutrinos  will contribute to  the relic  density, we must therefore
multiply~(\(\ref{eq:RelicDensity}\)) by 3  to obtain the final relic
DM abundance.

\subsection{Numerical Results}

The numerical analysis is separated  in two parts: in the first part, we perform
a scan  over the mSUGRA parameter  space to  calculate the
supersymmetric particle  spectrum and identify regions  where the LRHS
can be a possible candidate for CDM.  In the second part, we specify two mSUGRA  scenarios and calculate the  constraints on the
effective sneutrino annihilation coupling \(\lambda\rho\) by requiring
a sneutrino relic density of \(\Omega_{\rm DM} h^2=0.11\).

\begin{figure}[t]
\centering
\includegraphics[clip,width=0.49\textwidth]{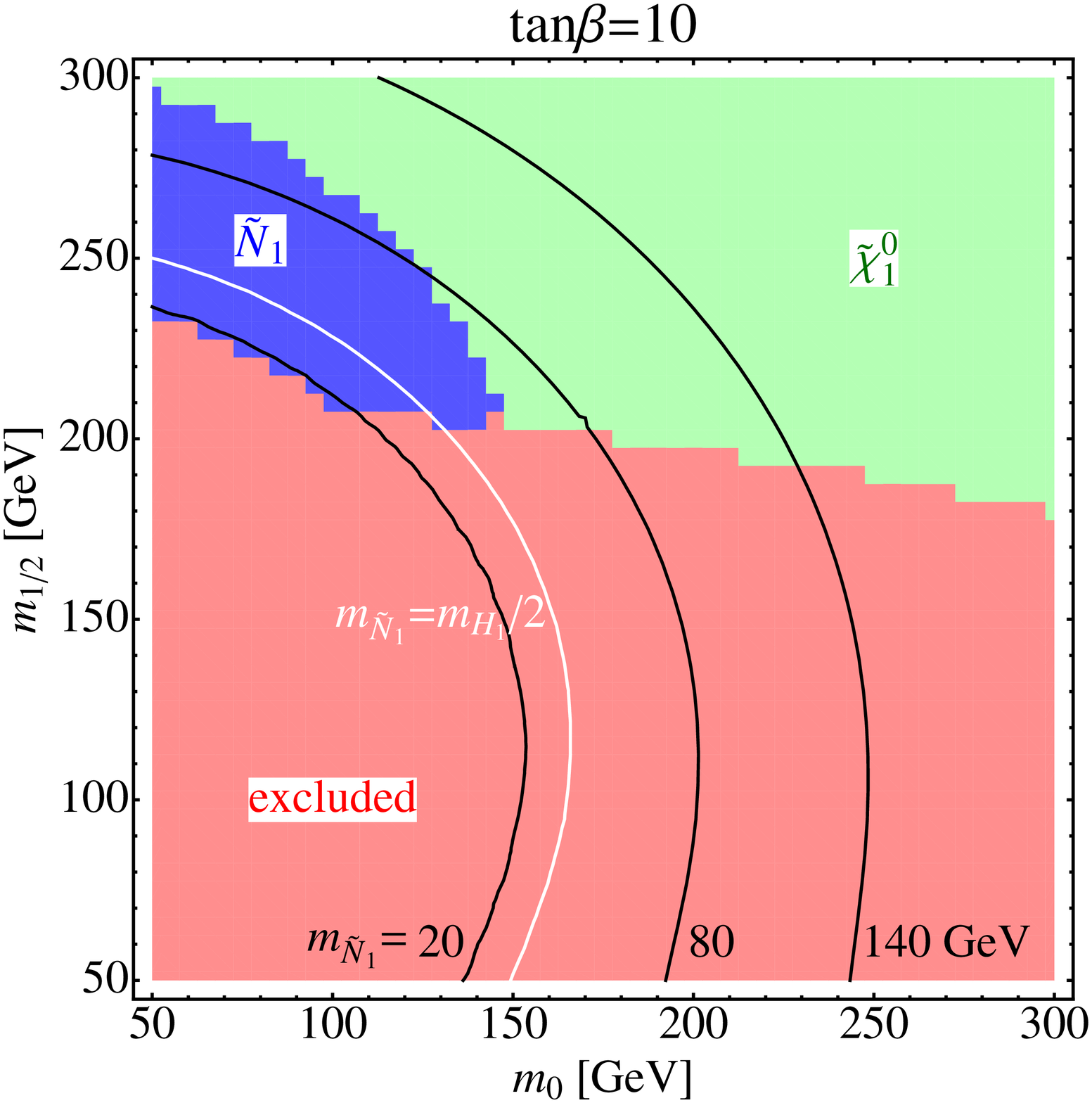}
\includegraphics[clip,width=0.49\textwidth]{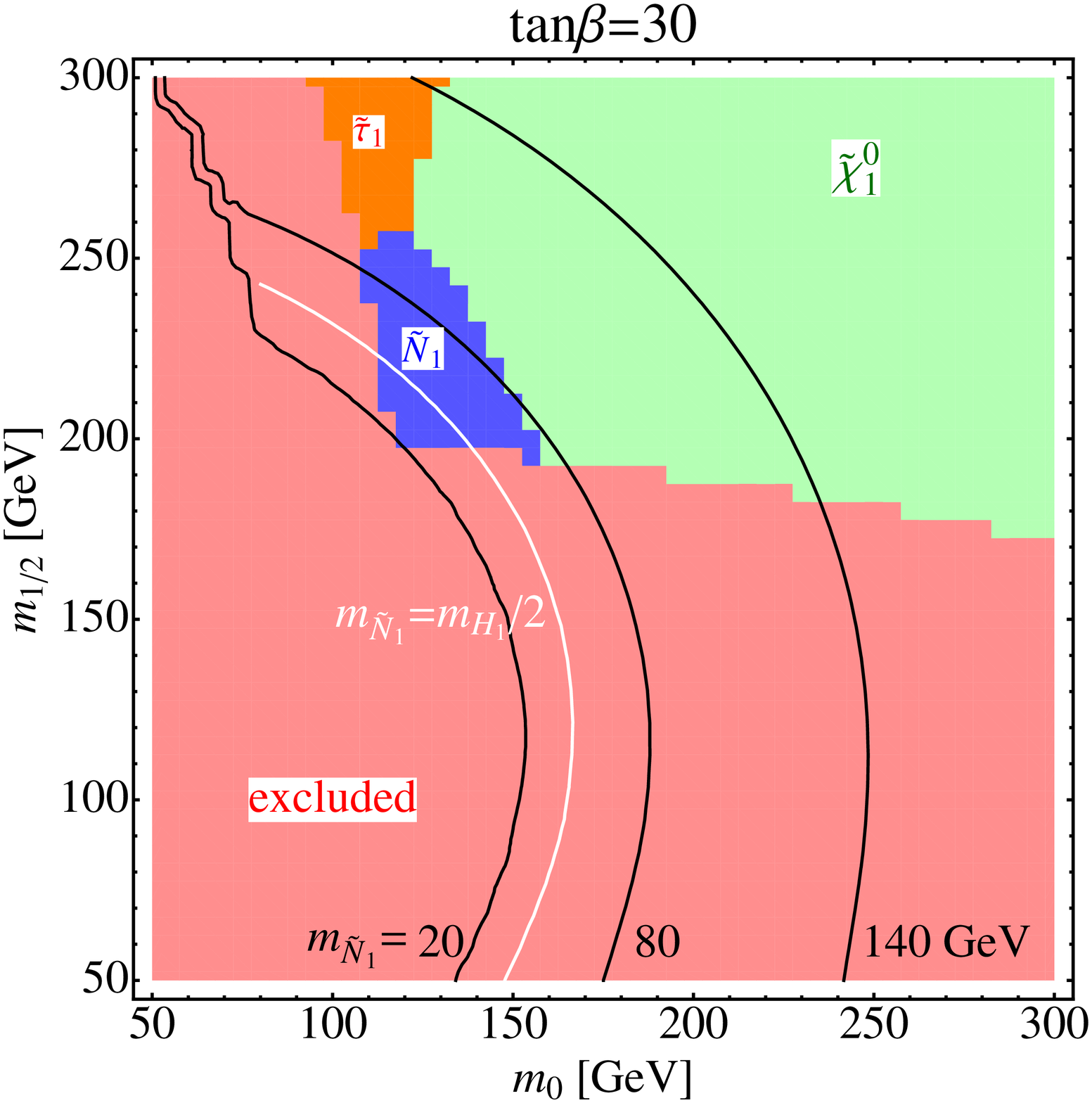}
\caption{\it Allowed  $(m_0,\,  m_{1/2})$  parameter  space for  a  mSUGRA
scenario  with  \(A_0=300\)~GeV,  \(\textrm{sign}\mu=+\), \(\lambda  =
\rho  = 10^{-2}\)  and \(\tan\beta=10\)  (left panel)  and  30 (right
panel).  The black contours show  the predicted LRHS mass, while the
sneutrino \(\tilde N_1\)/neutralino \(\tilde\chi^0_1\)/stau \(\tilde\tau_1\) LSP is  given by the blue/green/orange area.
The  red area is  excluded by  direct SUSY  mass searches.   The white
contour  is  defined by  the  condition \(m_{\tilde  N_1}=m_{H_1}/2\),
allowing  for   rapid  sneutrino  annihilation   via  the  $H_1$-boson
resonance.}\label{fig:scans}
\end{figure}

In  Figure~\ref{fig:scans}   we  plot  the   lightest  sneutrino  mass
\(m_{\tilde  N_1}\)   as  contours  in  the   mSUGRA  parameter  plane
(\(m_0,m_{1/2}\)), for two different  values of \(\tan\beta = 10\) (left)
and  30  (right).  In both plots of Figure~\ref{fig:scans}, we set \(A_0=300\)~GeV  and  \(\mu>0\). For  the
inflaton   couplings  \(\lambda,\rho\)   required  to   calculate  the
sneutrino masses~(\ref{eq:SneutrinoMasses}), we simply choose
\begin{equation}
	\lambda=\rho=10^{-2},
\end{equation}
in accordance with the bounds (\ref{Ulambda}) derived from inflation.
The  coloured areas in  Figure~\ref{fig:scans} denote  the LSP  in the
given  parameter region: sneutrino  \(\tilde N_1\)  (blue), neutralino
\(\tilde\chi^0_1\) (green) or  stau \(\tilde\tau_1\) (orange). The red
area on the bottom/left is  excluded by direct searches for SUSY particles. Specifically, the following experimental mass limits are used~\cite{PDG08}:
\begin{eqnarray} 
	m_{\tilde\chi^-_1} \!&>&\! 104\textrm{ GeV}\; , \nonumber\\
	m_{\tilde q}       \!&>&\! 375\textrm{ GeV}\; , \nonumber\\
	m_{\tilde g}       \!&>&\! 289\textrm{ GeV}\; , \\
	m_{\tilde \ell}    \!&>&\!  95\textrm{ GeV}\; , \nonumber\\
	m_{\tilde\nu_L}      \!&>&\! 130\textrm{ GeV}\; . \nonumber
\end{eqnarray}
Figure~\ref{fig:scans} was determined by appropriately using universal
soft SUSY-breaking parameters at the GUT scale according to the mSUGRA
scheme and then solving the  MSSM RG equations down to the electroweak
scale.  In  this respect,  our computation was  aided by  the software
package SPheno \cite{Porod:2003um}.  We  neglect the RG running of the
sneutrino parameters \(M_{\tilde N}^2\) and \(A_\rho\) which enter the
sneutrino  mass  matrix~(\ref{eq:SneutrinoMassMatrix}),  and  identify
them  directly with \(m_0^2\)  and \(A_0\),  respectively.  This  is a
reasonable approximation as  their RG evolution is only  driven by the
small couplings \(\lambda\) and \(\rho\). The Higgs coupling parameter
\(\mu\)   is  then   calculated  consistently   by   requiring  proper
electroweak symmetry breaking. In  the \(F_D\)-term model, the \(\mu\)
term  originates  from  the  VEV  of the  inflaton  (\ref{mu}).   This
immediately allows us to calculate  both the inflaton VEV, \(\langle S
\rangle=\mu/\lambda\),  and   the  mass  scale   of  the  right-handed
neutrinos,         \(m_N=\rho\langle        S         \rangle        =
\frac{\rho}{\lambda}\mu\)~(\ref{Wmodel}). For the \(\tilde
N_1\) LSP region of interest and with our choice \(\lambda=\rho=10^{-2}\), \(\mu\)
and  \(m_N\)  are  equal  and  of  order  300~GeV.   The  mass \(m_{\tilde N_1}\) of the LRHS as  LSP  ranges  between  20--100 GeV.  This 
allows  for a rapid  annihilation of \(\tilde N_1\) via  the Higgs  resonance, \(m_{\tilde
N_1}=m_{H_1}/2\approx   57\)~GeV,   along   the   white   contour   in
Figure~\ref{fig:scans}.

The \(F_D\)-term model puts strong constraints on the mSUGRA parameter
space, when requiring  a sneutrino LSP and taking  into account bounds
from  inflation.    As can be seen  in   Figure~\ref{fig:scans},  the
connection between LRHS mass \(\tilde N_1\) and \(\mu\) generally points towards
a low-energy SUSY  spectrum. This coincidentally includes the \(H_1\)-boson funnel region, where \(m_{H_1}\approx 2m_{\tilde
N_1}\).  On the
other  hand,  very large and small values  for  \(A_0\)  and \(\tan\beta\)  are
disfavoured  as  they  generally   exclude  a  sneutrino  LSP.   The above correlations may  be somewhat relaxed if non-universal
inflaton couplings \(\lambda\) and \(\rho\) are considered.

\begin{figure}[t]
\centering
\includegraphics[clip,width=0.90\textwidth]{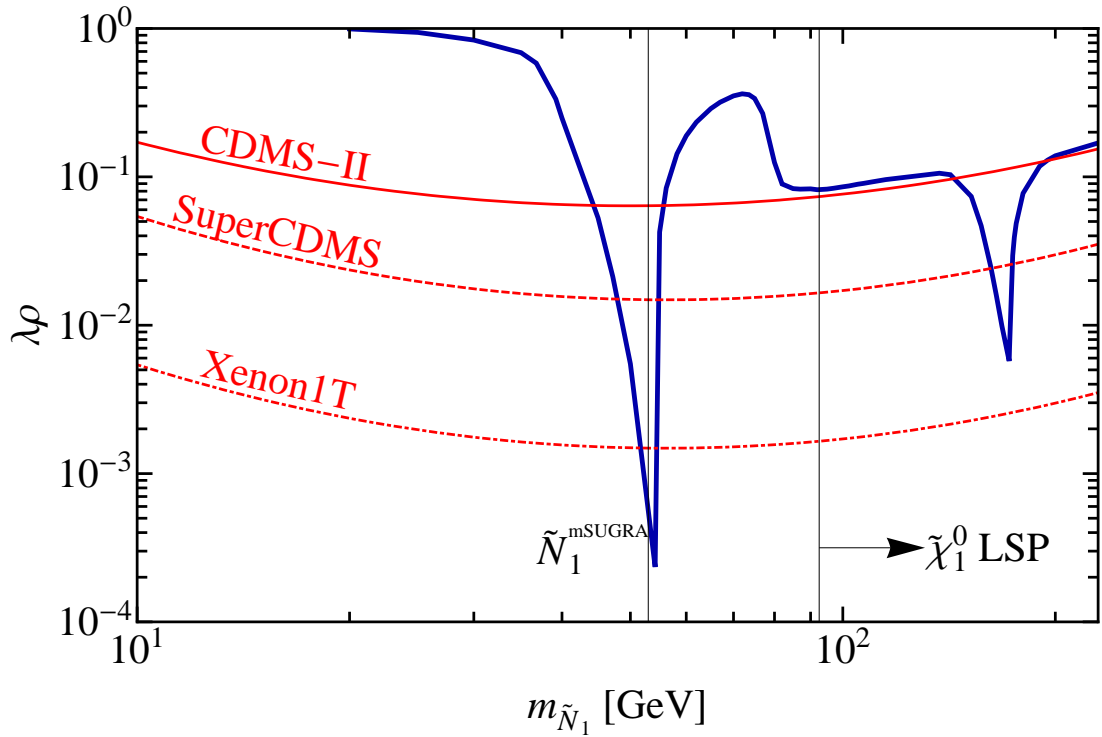}
\caption{\it  Effective  annihilation  coupling \(\lambda\rho\)  as  a
function of the  mass of the LRHS \(m_{\tilde  N_1}\) for the observed
relic density \(\Omega_{\rm DM}  h^2=0.11\) (blue curve) in the mSUGRA
Scenario~I~(\ref{ScenI}).  The actual  sneutrino and neutralino masses in the scenario
are indicated by vertical lines.  The red curves denote the upper bound on
\(\lambda\rho\) as obtained by the  CDMS-II experiment and as expected by
the projected sensitivities of SuperCDMS and Xenon1T.}
\label{fig:CDMscenarioI}
\end{figure}
\begin{figure}[t]
\centering
\includegraphics[clip,width=0.90\textwidth]{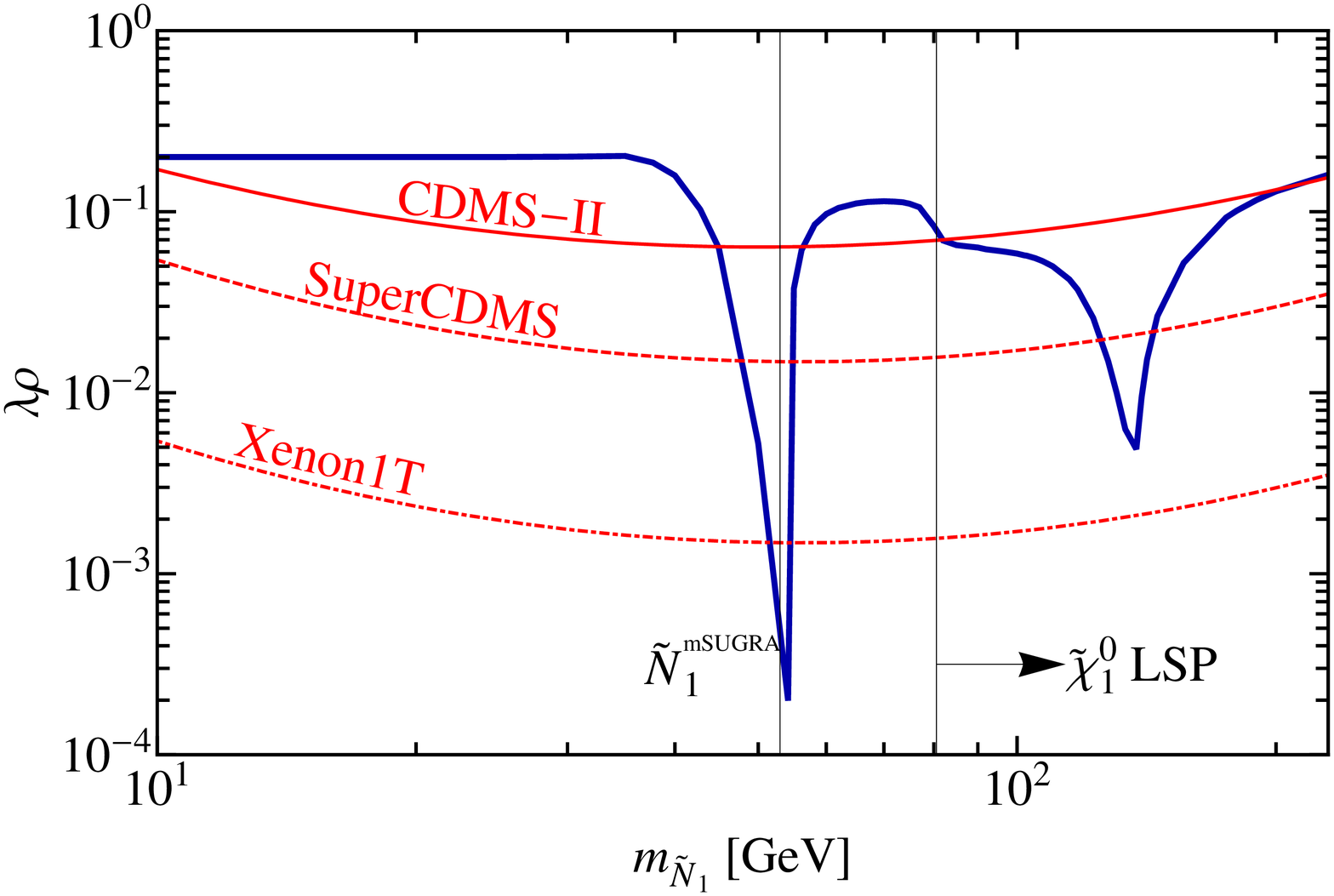}
\caption{\it As  in Figure~\ref{fig:CDMscenarioI}, but  for the mSUGRA
Scenario II~(\ref{ScenII}).}
\label{fig:CDMscenarioII}
\end{figure}

In  order to  compute  the  sneutrino relic  density  and analyze  the
constraints  on the  effective annihilation  coupling \(\lambda\rho\),
the following two mSUGRA scenarios have been selected:

\begin{itemize}

\item Scenario I:
\begin{equation}
  \label{ScenI}
m_0=70\textrm{ GeV},\ m_{1/2}=243\textrm{ GeV},\
A_0=300\textrm{ GeV},\ \tan\beta=10,\ \mu=303\textrm{ GeV}\; .
\end{equation}

\item Scenario II: 
\begin{equation}
  \label{ScenII}
m_0=125\textrm{ GeV},\ m_{1/2}=212\textrm{ GeV}, 
A_0=300\textrm{ GeV},\ \tan\beta=30,\ \mu=263\textrm{ GeV}\; .
\end{equation}

\end{itemize} 

In  addition, we  keep the  LRHS  mass as  a free  parameter. The  effective annihilation  coupling
\(\lambda\rho\)  (\ref{Llsp}) is then consistently calculated so as to obtain  a  sneutrino relic  density
\(\Omega_{\rm DM}\, h^2=0.11\), consistent with observation.  Furthermore,  we assume that the mass
splitting  between  the light  and  heavy  right-handed sneutrinos  is
sufficiently large so that co-annihilation can be safely ignored. This
is valid as  long as there is a sizeable mixing  between the light and
heavy right-handed  sneutrino states, which is certainly  true for the
mass  range \(m_{\tilde  N_1}< m_{\tilde\chi^0_1}\)  of  our interest.
All other MSSM parameters and masses were calculated within the mSUGRA
framework.  Numerical  estimates  of  the allowed  parameters  in  the
$(m_{\tilde{N}_1},\lambda\rho)$-plane are shown for Scenarios~I and~II
in  Figures~\ref{fig:CDMscenarioI} and Figure~\ref{fig:CDMscenarioII},
respectively.

As  we  have  seen  in Section~\ref{FDmodel}.3,  the  requirement  for
successful inflation puts upper  limits on the couplings $\lambda$ and
$\rho$. Given~(\ref{mSUGRA}),  (\ref{nmSUGRA}) and \ref{Ulambda}), the
upper limits on the product  $\lambda\rho$ for an inflaton sector with
a minimal and next-to-minimal K\"ahler potential may easily be deduced
to be
\begin{equation}
  \label{Inflimits}
\lambda\rho\ \stackrel{<}{{}_\sim}\ 2.3\times 10^{-4}\,,\qquad
\lambda\rho\ \stackrel{<}{{}_\sim}\ 5.8\times 10^{-4}\,,
\end{equation}
respectively, at the soft  SUSY-breaking scale $M_{\rm SUSY}$.  On the
other hand, Figures~\ref{fig:CDMscenarioI} and~\ref{fig:CDMscenarioII}
show that it should be
\begin{equation}
  \label{CDMlimits}
\lambda\rho\ \stackrel{>}{{}_\sim}\ 2\times 10^{-4}\; ,
\end{equation}
in  order  to account  for  the observed  DM  relic  abundance in  the
$H_1$-boson funnel region,  where \(m_{\tilde N_1}\approx m_{H_1}/2\). Larger
values of \(\tan\beta\)  do suppress the coupling required  to get the
observed  relic  density,  but  not  to a  level  compatible  with  the
inflationary constraints~(\ref{Inflimits}).  In  general, we find that
LRHS masses larger than about 100~GeV are not possible within a mSUGRA
realization of the $F_D$-term model.  This is indicated by the
value of the neutralino mass in the given mSUGRA scenario as displayed by vertical lines in Figures~\ref{fig:CDMscenarioI} and \ref{fig:CDMscenarioII}.

Further  constraints on the  $(m_{\tilde{N}_1},\lambda\rho)$-plane may
be obtained by taking into  account the limits from direct searches of
experiments  which  look for  scattering  between Weakly  Interacting
Massive Particles (WIMPs) and nuclei. Specifically, a WIMP, such as the
LRHS, can directly be detected through its elastic scattering with a
nucleus. In our case, the  relevant scattering process is \(\tilde N_1
+{}^A_Z X \to  \tilde N_1 + {}^A_Z X\) and  proceeds via a Higgs-boson
$t$-channel  exchange.   Its   cross-section  may  well  be  estimated
by~\cite{pospelov}
\begin{equation}
	\sigma_{\rm el}^{\rm nucleus}\ \approx\
	\frac{(1/2\lambda\rho)^2 v^2 |M_X|^2}{\pi}\
	\frac{m_{\rm red}^2}{m_{\tilde N_1}^2 m_{H_1}^4}\ ,
\end{equation}
where $m_{\rm  red}$ is the  reduced mass of the  LRHS-nucleus system,
i.e.
\begin{equation}
	m_{\rm red}\ =\
	\frac{m_{\tilde N_1}m_{X}}{m_{\tilde N_1}+m_X},
\end{equation}
and \(M_X\) is the nuclear matrix element. For comparison purposes, we
express  our results  in terms  of the  \emph{nucleon}  cross section.
Assuming the nucleus to be composed of \(A\) independent nucleons, the
nuclear  cross  sections  then  simply scale  quadratically  with  the
nucleon  number  \(A\)  and  the reduced  masses:  \(m^2_{\rm  red}(p)
\sigma_{\rm   el}^{\rm  nucleus}   =  A^2   m^2_{\rm   red}({}^A_Z  X)
\sigma_{\rm el}^{\rm nucleon}\).   The nucleon matrix element \(M_{\rm
nucleon}  \sim  10^{-3}\) is  mostly  sensitive  to the  strange-quark
Yukawa coupling.  An adequate estimate of the elastic scattering cross
section  $\sigma_{\rm el}^{\rm nucleon}$  of a  right-handed sneutrino
with a nucleon yields~\cite{pospelov}
\begin{equation}
  \label{eq:elastic}
	\sigma_{\rm el}^{\rm nucleon}\ \approx\
\left(5\times 10^{-50}~\textrm{ cm}^2\right)\,
	\left(\frac{\lambda\rho}{10^{-4}}\right)^2\,
	\left(\frac{100\textrm{ GeV}}{m_{H_1}}\right)^4\,
	\left(\frac{ 50\textrm{ GeV}}{m_{\tilde N_1}}\right)^2\; .
\end{equation}
The  upper limits  on  \(\lambda\rho\) are  derived  by comparing  the
estimate~(\ref{eq:elastic})   with   the    current   bound   on   the
spin-independent nucleon cross section from the CDMS-II experiment and
the       expected  sensitivities    of       the      SuperCDMS
extension~\cite{Ogburn:2006hk} and the Xenon1T experiment~\cite{Aprile:2006nz}.    These   limits   are  included   in
Figures~\ref{fig:CDMscenarioI}    and~\ref{fig:CDMscenarioII}.     The
current    bound    already    excludes    large    parts    of    the
$(m_{\tilde{N}_1},\lambda\rho)$-parameter   plane,   except   of   the
Higgs-boson  funnel  regions.   In   the  near  future,  the  upgraded
experiment SuperCDMS  will cover a large part of the parameter space,  but it
will  leave  open  the  lightest  Higgs-boson  pole  region  which  is
theoretically favoured by inflation  within the mSUGRA framework.  The
proposed   Xenon1T  experiment is   expected  to
further narrow  down this uncovered parameter range  of the $F_D$-term
model.

Dark Matter may also be indirectly searched for through the detection of its final annihilation products, such as photons, positrons, anti-protons or neutrinos. The dominant channel of the LRHS annihilation in the Higgs funnel is determined by an effective scalar coupling with a $b\bar b$ pair, which is approximately independent of the relative velocity of the annihiliating sneutrinos. Rates at low temperatures resulting in gamma-ray or charged particle fluxes are therefore not suppressed compared to the rates at the freeze-out temperature responsible for the LRHS relic density. There are several signals that could be explained as an observation of DM annihilation but, as of now, do not provide a consistent picture interpretable by a single DM candidate and model. For example, the excess in the diffuse galactic gamma ray spectrum measured by the EGRET detector may be interpreted by a 50-100~GeV WIMP, as given by the LRHS in our model, whereas the 511~keV line observed by the INTEGRAL satellite would hint at an MeV DM particle~\cite{Taoso:2007qk,Hooper:2007vy}. Upcoming projects such as the GLAST and PAMELA satellites will have higher sensitivities, probe new energy ranges and should provide a clarification of the observational status.  
High-energy neutrinos as annihilation products are expected and can be searched for in the Sun and the Earth, as WIMPs can accumulate in their centre. For the LHRS there is no spin-dependent coupling to nuclei, and its capture rate along with the produced neutrino flux is suppressed. In addition, for an annihilation via the Higgs resonance, the effective annihilation coupling required to get the correct relic density is very small. The LRHS is therefore not expected to be within the reach of high-energy neutrino telescopes~\cite{Beltran:2008xg}, such as IceCube~\cite{Rott:2007zz}.

\bigskip

\setcounter{equation}{0}
\section{Conclusions}\label{conclusions}

We  have  analyzed in  detail  the  relic  abundance of  the  lightest
right-handed sneutrinos (LRHS)  in the supersymmetric $F_D$-term model
of hybrid inflation.  The  inflationary potential of the model results
from  the  $F$-term  of  the inflaton  multiplet  $\widehat{S}$.   The
$F_D$-term  model also includes  a subdominant  non-anomalous $D$-term
generated from  the local U(1)$_X$  symmetry of the  waterfall sector,
which does not affect the  inflaton dynamics.  As was mentioned in the
introduction and further discussed in Section~\ref{FDmodel}, the model
adequately  fits the  current CMB  data  of inflation  and provides  a
natural  solution to  the so-called  gravitino  overabundance problem,
without   resorting   to   an   excessive  suppression   of   possible
renormalizable  couplings  of  the  inflaton to  the  MSSM  particles.
Finally, the  $F_D$-term model closely relates  the $\mu$-parameter of
the MSSM to an SO(3) symmetric  Majorana mass $m_N$ through the VEV of
the inflaton  field.  If $\lambda  \sim \rho$, this implies  that $\mu
\sim m_N$, so the  model may naturally predict lepton-number violation
at  the electroweak  scale and  potentially  account for  the BAU  via
thermal resonant leptogenesis.

In spite of the  explicit lepton-number violation through the Majorana
term  $\frac{1}{2}\, \rho\, \widehat{S}  \widehat{N}_i \widehat{N}_i$,
the $F_D$-term  hybrid model conserves  $R$-parity.  Consequently, the
LSP of the spectrum is stable and so qualifies as candidate to address
the CDM problem. The new aspect of the $F_D$-term hybrid model is that
thermal right-handed sneutrinos emerge as new candidates to solve this
problem,  by  virtue of  the  quartic coupling:  $\frac{1}{2}\,\lambda
\rho\,  \widetilde{N}^*_i \widetilde{N}^*_i  H_u H_d\  +\  {\rm H.c.}$
This  new quartic  coupling results  in the  Higgs potential  from the
$F$-terms of  the inflaton field,  and it is  not present in  the more
often-discussed  extension of  the MSSM,  where  right-handed neutrino
superfields have  bare Majorana  masses. Provided that  the couplings
$\lambda$  and  $\rho$  are   not  too  small,  e.g.~$\lambda,\,  \rho
\stackrel{>}{{}_\sim} 10^{-2}$, the  LRHS $\widetilde{N}_{\rm
LSP}$ as LSP can   efficiently  annihilate  via   the  lightest  Higgs-boson
resonance  $H_1$ into  pairs of  $b$-quarks, in  the  kinematic region
$m_{H_1}  \approx 2 m_{\widetilde{N}_{\rm  LSP}}$, and  so drastically
reduce its relic density  to the observed value: $\Omega_{\rm DM}\,h^2
\approx 0.11$. 

Experiments, such  as CDMS-II, SuperCDMS  and Xenon1T, which  look for
signatures  of WIMPs  through their  elastic scattering  with nuclei,
will  significantly  constrain  the  allowed parameter  space  of  the
$F_D$-term  model. They  will  exclude most  of  the parameter  space,
except possibly of  a narrow region close to  the lightest $H_1$-boson
resonance, where $m_{H_1}  \approx 2 m_{\widetilde{N}_{\rm LSP}}$.  It
might seem that to obtain  this particular relation between the masses
of the $H_1$ boson and $\widetilde{N}_{\rm LSP}$, a severe tuning of the
model parameters is required. However, it is worth stressing here that
such a mass relation may  easily be achieved within a mSUGRA framework
of the $F_D$-term model that successfully realizes hybrid inflation.

The  LRHS scenario  of  CDM requires  relatively  large $\lambda$  and
$\rho$  couplings that could,  in principle,  make Higgs  bosons decay
invisibly,   e.g.~$H\to\widetilde{N}_{\rm   LSP}\,  \widetilde{N}_{\rm
LSP}$.  Also, right-handed sneutrinos  could be present in the cascade
decays  of   the  heavier  supersymmetric   particles.   The  collider
phenomenology  of such a  CDM scenario  lies beyond  the scope  of the
present  article.   Instead,  we   note  that  the  $F_D$-term  hybrid
inflationary model  can give rise  to rich phenomenology which  can be
probed at  high-energy colliders~\cite{NprodLHC,NprodILC}, as  well as
in low-energy experiments of lepton flavour and number violation, such
as  $0\nu\beta\beta$ decay,  $\mu\to e\gamma$~\cite{CL},  $\mu\to eee$
and $\mu\to e$ conversion in nuclei~\cite{IP,LFVN}. It would therefore
be  very interesting to  systematically analyze  possible correlations
between predictions for  cosmological and phenomenological observables
in the $F_D$-term model.

\subsection*{Acknowledgements}
This work is supported in part by the PPARC research grant: PP/D000157/1.

\end{document}